\documentclass[preprint,12pt]{elsarticle}

\usepackage{amssymb}

\newcommand{\gsim}{\raisebox{-0.13cm}{~\shortstack{$>$ \\[-0.07cm] $\sim$}}~}
\begin{document}

\begin{frontmatter}


 \author{Keiko~I.~Nagao\corref{cor1}}
\ead{knagao@post.kek.jp}
 \address{KEK Theory Center, IPNS, KEK, 1-1 Oho, Tsukuba, 305-0801, Japan}
   \author{Tatsuhiro~Naka\corref{cor2}}
\ead{naka@flab.phys.nagoya-u.ac.jp}
 \address{Institute for Advanced Reseach, Nagoya University, Nagoya 464-8602, Japan}

\title{Isospin Violating Dark Matter Search by Nuclear Emulsion Detector}


\author{}

\address{}

\begin{abstract}
Dark matter signal and its annual modulation of event number are observed by some direct searches.
However, the parameter spaces have been excluded by other experiments. 
The isospin-violating dark matter is a hopeful candidate to solve the discrepancy. 
We study the possibility that a future project of dark matter search  using nuclear emulsion can reach the favored region by the isospin-violating dark matter.
Since the detector has the directional sensitivity, it is expected to examine the region including the modulation property.
\end{abstract}

\begin{keyword}
dark matter, direct detection, nuclear emulsion, isospin-violating dark matter
\end{keyword}

\end{frontmatter}


\section{Introduction}
Dark Matter is a notable subject in both particle physics and astrophysics. A number of observations and experiments have been performed in order to survey its nature. 
Wilkinson Microwave Anisotropy Probe (WMAP) \cite{Komatsu:2010fb} has revealed that dark matter consists about $23\%$ of the energy density of the Universe. 
On the other hand, the direct and indirect detection of dark matter also put constraints on the interaction of dark matter \cite{DM_rev}. In the paper, we focus on the direct detection experiments. DAMA \cite{Bernabei:2000qi,Bernabei:2008yi}, CoGenT \cite{CoGenT2011} and CRESSTII \cite{Angloher:2011uu} presented data which can be interpreted as dark matter signal. The positive data suggests light dark matter mass $\sim O(10)$GeV. 
However, the parameter space have been excluded by CDMS-II \cite{Ahmed:2010wy}, XENON10 \cite{Angle:2011th} and XENON100 \cite{Aprile:2011hi}.
There are many attempts to explain the discrepancy, for example, dark matter nature such as the inelastic dark matter \cite{TuckerSmith:2001hy, Kopp:2009qt, Chang:2010yk}, atomic uncertainties such as scintillation efficiency factor \cite{Savage:2010tg}, quenching factor \cite{Hooper:2010uy} and channeling fraction \cite{Drobyshevski:2007zj,Bernabei:2007hw,Savage:2008er,Bozorgnia:2010xy}, and astrophysical uncertainties \cite{Frandsen:2011gi}.

Isospin-Violating Dark Matter (IVDM) can solve the contradiction supposing specific dark matter nature. 
The coupling between dark matter and proton in nuclei $f_p$ and that between dark matter and neutron $f_n$ are usually supposed to be same, i.e. $f_n/f_p=1$. If one take the ratio not equal to one but $f_n/f_p \approx -0.7$, part of DAMA and CoGent signal region is allowed by other experiments \cite{Feng:2011vu}. Many studies on IVDM have been done including from aspects of indirect detection and collider constraints \cite{Feng:2011vu,Frandsen:2011ts,Schwetz:2011xm,Farina:2011pw,McCabe:2011sr,Kelso:2011gd,Chen:2011vda,Gao:2011bq,Kumar:2011dr}. The origins of isospin violation are also proposed from the particle physics point of view \cite{Khlopov:2008ki,Kang:2011wb,DelNobile:2011je,Frandsen:2011cg,Gao:2011ka,Cline:2011zr,Kawase:2011az,DelNobile:2011yb,He:2011gc}.

A nuclear emulsion is a type of photographic film; it can detect the tracks of charged
particles emitted by the dark matter-nucleus scattering. Therefore it has the directional 
sensitivity in dark matter search. The mechanism is as follows.
Silver halide crystals densely dispersed in gelatin are penetrated by charged
particles, and they become visible silver grains via development treatment. The
detector is used in experiment such as neutrino oscillation using a lot of nuclear emulsion
films $\sim$30000 kg \cite{Eskut:2007rn,Kodama:2007aa,Agafonova:2010dc}.
Spatial resolution of the nuclear emulsion is defined by crystal size and number density per volume. 
The nuclear emulsion detector has a advantage in the directional dark matter search since its spatial resolution is extremely higher than other detectors which has directional sensitivity \cite{Grignon:2009zc,Miuchi:2010hn,Ahlen:2010ub,Vahsen:2011qx}. Large mass as a solid detector, which enables to achieve high sensitivity, is another merit of the nuclear emulsion. Status of the experiment is in research and development (R\&D) now.

Since the nuclear detector has the directional sensitivity, it can test the annual modulation of dark matter including directional information. Therefore, testing favored region by DAMA, CoGenT and CRESST is one of the attractive possibility of the detector. We study the sensitivity of the nuclear emulsion detector for IVDM, and discuss the possibility that it test the favored region of IVDM.

This article is organized as follows. In Sec.\ref{sec:emulsion}, we introduce the concept and expected sensitivity of the direct detection experiment with nuclear emulsion. In Sec.\ref{sec:result}, we introduce the IVDM and show the sensitivity of the emulsion detector for it. We conclude in Sec.\ref{sec:summary}.

\section{Direct Detection with Nuclear Emulsion}
\label{sec:emulsion}
In dark matter detection with nuclear emulsion, the recoiled particle by dark matter-nucleus scattering is detected 
as track in nuclear emulsion. Hence the recoil energy of dark matter corresponds to the length of track.
$O(10)$GeV mass dark matter, whose typical energy is $O(10)$keV, leaves
submicron track on nuclear emulsion layers.
Detection of submicron track length had been difficult with ordinary nuclear emulsion whose spatial resolution is about 1 $\mu$m. However, 
it is finally achieved due to recent progress of nuclear emulsion technology, which enable to make fine-grained
nuclear emulsion \cite{Natsume:2007zz}.
Therefore currently it is possible to detect tracks of dark matter with $ O(10)$GeV mass.
This idea is very unique among the directional dark matter search using solid detector. 

Current components of the detector is summarized in Table\ref{Table:isotopes}.
Target nuclei are silver (Ag) and bromine (Br) as heavy targets. On the other hand, as light targets, carbon (C), nitrogen (N) and oxygen (O) are included in the emulsion layers. Current limit of detectable range is about 100 nm, which corresponds to the recoil energy 160 keV for
heavy targets and 33 keV for light targets\footnote{See Appendix A for detailed correspondence between track range and the recoil energy.}. 
The thresholds are expected to be improved by R\&D henceforth.

\begin{table}[tbp]
\hspace*{.15\textwidth}
\begin{minipage}{.6\textwidth}
\begin{center}
\begin{tabular}{|c||r|r|r|}
\hline
\multicolumn{1}{|c|}{\,} & \multicolumn{1}{c|}{Weight(\%)} & \multicolumn{2}{c|}{$A_i$(abundance)} \\ \hline\hline
Ag                                & 39.65                                          &107(51.84) &109(48.16)    \\ \hline
Br                                 & 29.01                                          &79(50.69)  &81(49.31)      \\ \hline
O                                  & 11.76                                          &16&                     \\ \hline
C                                  & 11.72                                          &12(98.9)    &13(1.1)          \\ \hline
N                                  & 4.57                                            &14&                   \\ \hline
H					&2.27&1 &\\ \hline
S					&0.05&32(95.02)&34(4.2)\\ \hline
I					&0.96&127&\\ \hline
\end{tabular}
\end{center}
\end{minipage}\\
\caption{Contents of nuclear emulsion layers. In the second and the third column, mass number of isotope $A_i$ (its natural abundance ratio) is shown. We omit isotopes whose abundance ratio is less than $1$\%. }
\label{Table:isotopes}
\end{table}

\begin{figure}[th]
\label{fig:emulsionsensitivity}
\begin{center}
\includegraphics[width=130mm]{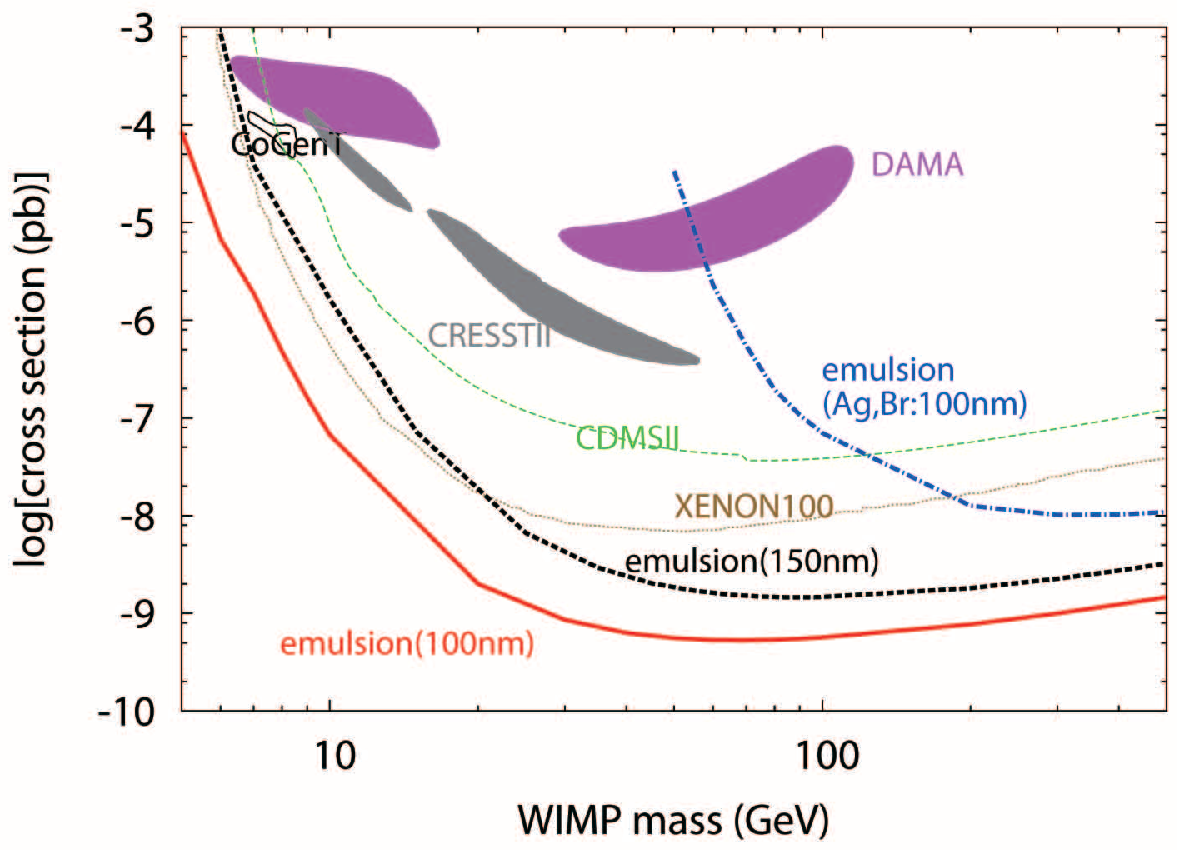}
\caption{Expected sensitivity of the nuclear emulsion detector and current results of direct searches. Regions filled with magenta (black in monochrome printing), light gray and white are positive signal by DAMA, CoGenT and CRESST, respectively. Thin dashed (dotted) line represents null constraint by CDMSII (XENON100). Thick solid and thick dashed lines correspond to the expected sensitivity of the nuclear emulsion detector with the detectable range thresholds 100nm and 150nm, respectively. We also show the sensitivity of the nuclear emulsion adopting target nuclei only heavy atoms (Ag and Br) as thick dot-dashed line.}
\end{center}
\end{figure}

The event rate of the direct search experiment $R$ is defined the integral of the recoil energy $E_R$ and the dark matter velocity in the frame of the Earth $v$, as
\begin{eqnarray}
R=N_T n_\chi \int_{E_{R,\mathrm{min}}} dE_R \int_{v_{\mathrm{min}}}^{v_{\mathrm{max}}} d^3 v \,\,f(v) v \frac{d\sigma_A}{dE_R}
\label{eq:eventrate}
\end{eqnarray}
where $N_T$ is the number of target nuclei, $n_\chi$ is the number density of dark matter, $f(v)$ is the distribution function of dark matter velocities,
and $\sigma_A$ is the cross section of the dark matter-nucleus elastic scattering \cite{Lewin:1995rx}. 
We assume a Maxwell-Boltzmann distribution:
\begin{eqnarray} 
f(v)=\frac{1}{(\pi v_0^2)^{3/2}}e^{-(v+v_E)^2/v_0^2},
\end{eqnarray}
and use $v_0=220$ km/sec as the velocity of the Solar-system relative to the galactic halo, and $v_E=230$ km/sec as the Earth's velocity relative to the dark matter distribution.
The annual modulation effect is omitted. 
In addition, local dark matter density $\rho_0= 0.3$ GeV$/$cm$^{3}$ and Helm form
factor \cite{Helm:1956zz} are used as fundamental parameters. 
Perfect detection efficiency and zero backgrounds are assumed. 

We show the 90\% C.L. sensitivity of the nuclear emulsion detector for dark matter and nucleon cross section
of spin-independent interaction with 1000 kg$\cdot$year exposure in
Fig.1. 
Signal regions by DAMA \cite{Savage:2008er}, CoGenT \cite{CoGenT2011} and CRESSTII \cite{Angloher:2011uu}, and constraints by CDMSII \cite{Ahmed:2010wy} and XENON100 \cite{Aprile:2011hi} are also presented for references. Thick solid line with ``100nm'' label (thick dashed one with ``150nm") corresponds the sensitivity of nuclear emulsion detector for cases the detectable range threshold is 100nm (150nm), namely, the energy threshold $\sim 33$ keV ($\sim 44$ keV). Current limit of the detectable range  is 100nm, therefore the blue line with ``150nm'' label is fairly conservative. 
For comparison, the sensitivity of nuclear emulsion detector adopting only Ag and Br for targets, is represented as thick dot-dashed line while thick solid and dashed ones correspond to the case adopting Ag, Br, C, N, and O as target nuclei. We note that good sensitivity in light dark matter mass region $\sim 10$ GeV can be achieved only when the effect of the dark matter-light nuclei (C, N, and O) scattering is included. 

In realistic experiment, nuclear emulsion detector will be mounted on equatorial telescope
and put toward to expected direction of incoming dark matter (i.e., direction of Cygnus in the celestial sphere) in
underground facility because the detector does not have time resolution. 
The detector should
be constructed in underground and in clean room to suppress background signals.
Readout of signal tracks will be done by optical microscope and X-ray microscope.
First, candidate tracks are searched automatically by optical microscopy \cite{Kimura2012}, and finally they are
confirmed by X-ray microscope which has higher resolution than optical one. These
new technologies have already confirmed in \cite{Naka2012}. Ultimately, angular resolution is expected
to be about 15-20 degree for X-ray microscope observation.

\section{Iso-Spin Violating Dark Matter in Emulsion Dark Matter Search}
\label{sec:result}
First we shortly review the IVDM for the simplest case that the target contains only one type of atom which one isotope dominates.
The differential cross section of spin-independent scattering in Eq.(\ref{eq:eventrate}) can be represented as $d\sigma_A/dE_R=\sigma_A m_A/(2v^2 \mu_A^2)$ where $m_A$ is the nucleus mass and  
$\mu_A$ is the reduced mass defined with dark matter mass $m_\chi$ as $\mu_A=m_Am_\chi/(m_A+m_\chi)$.
 The dark matter-nucleus scattering cross section can be described as
\begin{eqnarray}
{\sigma}_A = \frac{\mu_A^2}{\Lambda^4}\left[f_p Z F_A^p(E_R) +f_n(A-Z) F_A^n(E_R)\right]^2 
\label{eq:sigmaA}
\end{eqnarray}
where $\Lambda$, $Z$ and $F_A^{p(n)}(E_R)$ are 
the scale which parametrizes the scattering, the atomic number of $A$, and the proton (neutron) form factor for nucleus $A$, respectively. 
$(A-Z)$ corresponds to the neutron number of nucleus $A$. 
Since the difference between $F_A^p(E_R)$ and $F_A^n(E_R)$ is negligible compared to that of couplings,  we set both of the form factors are same as $F_A(E_R)$ afterward.
Usually the dark matter-proton coupling is assumed to be same as the dark matter-neutron coupling, i.e. $f_n=f_p$. 
In that case, Eq.(\ref{eq:sigmaA}) is simplified as $\sigma_A \propto \mu_A^2 f_n^2 A^2$, which implies the well-known 
property that the spin-independent scattering cross section is propotional to the squared mass number. 
If there are more than one isotope, Eq.(\ref{eq:sigmaA}) becomes
\begin{eqnarray}
{\sigma}_A = \sum_i \eta_i \frac{\mu_{A_i}^2}{\Lambda^4} F_A(E_R)^2 \left[f_p Z +f_n(A_i-Z)\right]^2 
\label{eq:sigmaAisotopes}
\end{eqnarray}
where $\eta_i$ is the abundance ratio of isotope $A_i$.
We introduce the dark matter-proton scattering cross section $\sigma_p=f_p^2\mu_p^2F_A(E_R)^2/\Lambda^4$, then Eq.(\ref{eq:sigmaAisotopes}) can be written as
$\sigma_A=\sigma_p \sum_i \eta_i \mu_{A_i}^2/\mu_p^2 \left[Z +f_n/f_p(A_i-Z)\right]^2$. Hence the constraints for $\sigma_p$ varies with $f_n/f_p$. 
We introduce the dark matter-nuclei scattering cross section for $f_n/f_p=1$ case as $\sigma_N$, which is nothing but the cross section measured by experiments shown in Fig.1. 
Taking negative $f_n/f_p$, the constraints for $\sigma_p$ are suppressed compared to that for $\sigma_N$. 
Especially for $f_n/f_p=-0.7$, the null constraints of XENON10, 100 and CDMSII are woundy suppressed while the suppression for positive constraints is mild.  
As a consequence, part of positive signal region is allowed by null results for $m_\chi \sim 8$GeV and $\sigma_p \sim 2\times 10^{-2}$pb \cite{Feng:2011vu}. 

We extend Eq.(\ref{eq:sigmaAisotopes}) for the case that the target consists of several spices of atom, which is labeled by $j$. Therefore $A^j_i$ is the  isotope $i$ of atom $j$. 
Since not only $\sigma_A$ but also other parameters in the event rate $R$ depend on atom, we will start the discussion from the event rate again.
The number of target nuclei $N_T$ in unit of $/kg$\footnote{In the context of direct detection, the event rate is conventionally represented in unit of /kg/day. We adopt the notation and use /kg unit for $N_T$.} is expressed as $N_0\times 10^3/\tilde{A}$ where $N_0$ is the Avogadro number and $A$ is the Molar mass in $g/mol$ unit, that is nothing but the mass number.
The event rate of this case is 
\begin{eqnarray}
R\hspace{-.5em}&=&\hspace{-.5em}\sigma_p \sum_j \xi_j \left(\sum_i \frac{N_0\times 10^3}{A_i^j}  \eta_i \frac{m_{A^j_i}^2}{\mu_p^2} [Z_j+f_n/f_p(A^j_i-Z_j)]^2 n_\chi\right. \nonumber \\
&&\times \left. \int_{E_{R,\mathrm{min}}} \hspace{-1.2em} dE_R \int_{v_{\mathrm{min}}}^{v_{\mathrm{max}}} d^3 v \,\,f(v) v F_{A^j_i}(E_R)^2 \right)
\label{eq:eventrate_severalatoms}
\end{eqnarray}
where $\xi_j$ is the weight ratio of atom $A^j$. When we can suppose that the form factor $F_{A^j_i}(E_R)$ varies mildly for different $A^j_i$, and the thresholds of the recoil energy ${E_{R,\mathrm{min}}}$ are common for all target atoms, Eq.(\ref{eq:eventrate_severalatoms}) is written as
\begin{eqnarray}
R&=&\sigma_p \sum_j \left[ \xi_j \left(\sum_i \frac{N_0\times 10^3}{A_i^j}  \eta_i \frac{m_{A^j_i}^2}{\mu_p^2} [Z_j+f_n/f_p(A^j_i-Z_j)]^2\right)\right. \nonumber\\
&&\left. \times \left(n_\chi \int_{E_{R,\mathrm{min}}} dE_R \int_{v_{\mathrm{min}}}^{v_{\mathrm{max}}} d^3 v \,\,f(v) v F_{A^j_i}(E_R)^2 \right)\right],
\end{eqnarray}
and the second parenthesis is independent of $i$ and $j$. 
The ratio of $\sigma_N$ and $\sigma_p$ is described as
\begin{eqnarray}
\frac{\sigma_p}{\sigma_N}=\frac{\sum_j \xi_j \left(\sum_i \eta_i m_{A_i^j} A_i^{j} \right)}{\sum_j \xi_j \left(\sum_i \eta_i m_{A_i^j}/A_i^j \left[Z_j+f_n/f_p(A_i^j-Z_j)\right]^2\right)} .
\label{eq:ratio_severalatoms}
\end{eqnarray} 

\begin{figure}[th]
\begin{center}
\label{fig:IVDMemulsion}
\includegraphics{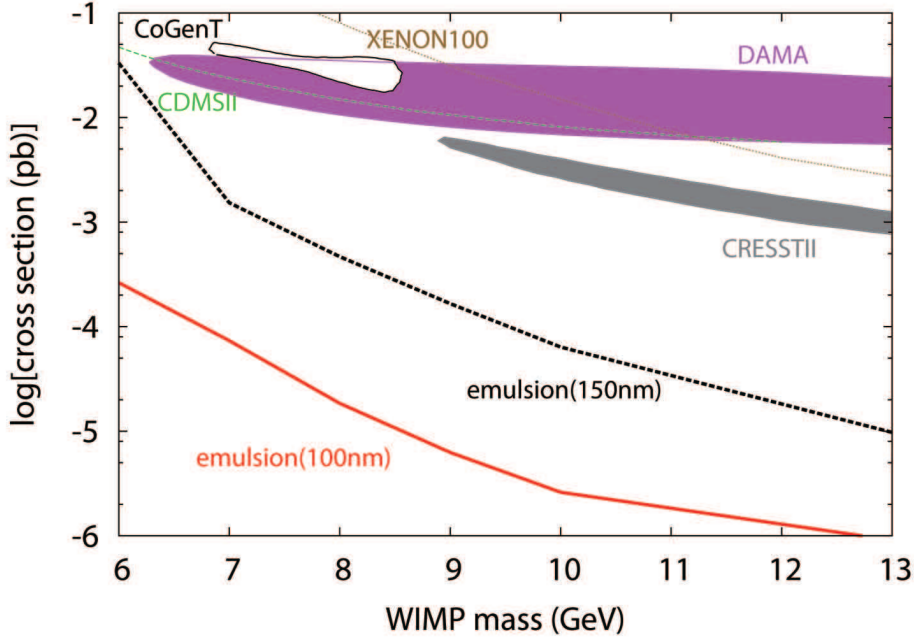}
\caption{Expected sensitivity of the nuclear emulsion detector to the IVDM for $f_n/f_p=-0.7$. Legend is same as Fig. 1.}
\end{center}
\end{figure}

Now we are ready to estimate the expected constraint by the nuclear emulsion detector for the IVDM. Target atoms and its isotopes in nuclear emulsion layers are listed in Table1. Ag and Br are dominant components, however, they are so heavy that they do not have sensitivity for light mass region as we comment in Section~2.
Instead, light atoms C, N and O are sensitive to light dark matter. In the calculation, we assume that the target atoms are only C, N and O and convert the constraints in Fig.1 to case of IVDM.
Since C, N, and O have almost same atomic number $A$ and the energy threshold $E_{R,\mathrm{min}}$ (See Appendix B for the energy thresholds of each atom), the supposition that mild variation of the form factor and the common threshold for target atoms, are justified. The result is shown in Fig.2. For reference, we also show the constraints for the IVDM by other experiments: signal regions of DAMA/LIBRA 3-$\sigma$ result \cite{Savage:2008er} with no-ion channeling effect, CoGenT 90\% C.L. \cite{CoGenT2011}, CRESST $2\sigma$ region \cite{Angloher:2011uu}, and null constrains of 
XENON100  90\%C.L. \cite{Aprile:2011hi} 
 and CDMSII 90\% C.L. \cite{Ahmed:2010wy}. 
We do not include uncertainties such as scintillation efficiency factor, quenching effect, and astrophysical ones.
Thick solid line (thick dashed line) represents the expected sensitivity of the nuclear emulsion detector with the range threshold $100$nm ($150$nm). 
Since the choice of $f_n/f_p$ is destructive, the cross section for $f_n/f_p=-0.7$ is suppressed compared to the isospin-conserving case. 
Especially if the number of neutrons in nucleus is much larger than that of protons, the suppression is amplified. 
Suppression for the nuclear emulsion is small because the number of neutron in  target atoms (C, N, and O) is almost same as that of proton.
Therefore the sensitivity of the nuclear emulsion detector for isospin-violating dark matter is enough to test the region favored by the IVDM. 

In order to examine the region favored by IVDM, the sensitivity for light dark matter is required. 
For nuclear emulsion,
such a small recoil energy corresponds to short length tracks near detector threshold. 
Therefore, rejection of the backgrounds with very short length will be important.
In Appendix B, we discuss the realization of good back ground rejection in detail. 

\section{Summary and Discussion}
\label{sec:summary}
{We have examined the expected sensitivity of the future detector using the nuclear emulsion. Especially we study the sensitivity for the IVDM. The region favored by IVDM, i.e. signal region by DAMA and CoGenT  where is not excluded by  XENON10, 100 and CDMSII, will be reached by the nuclear emulsion detector if it achieves good detection efficiency and back ground rejection. In that case, it can test the signal region of other experiments including the signal direction since the detector has the directional detectability.

\subsubsection*{Acknowledgment}
This work is supported by the Grant-in-Aid for Nagoya
University Global COE Program, ``Quest for Fundamental Principles in the
Universe: from Particles to the Solar System and the Cosmos",  
the Grant-in-Aid for Scientific Research on Innovative Areas (No.2310400) and 
 Grant-in-Aid for Research Activity Start-up (23840018) from the Ministry
of Education, Culture, Sports, Science and Technology of Japan (MEXT).
KN thanks M.~Nojiri and M.~Yamanaka for useful discussions. 

\section*{Appendix A: Track Length and Recoil Energy}
\begin{figure}[t]
\begin{center}
\label{fig:correspondence}
\includegraphics[scale=.8]{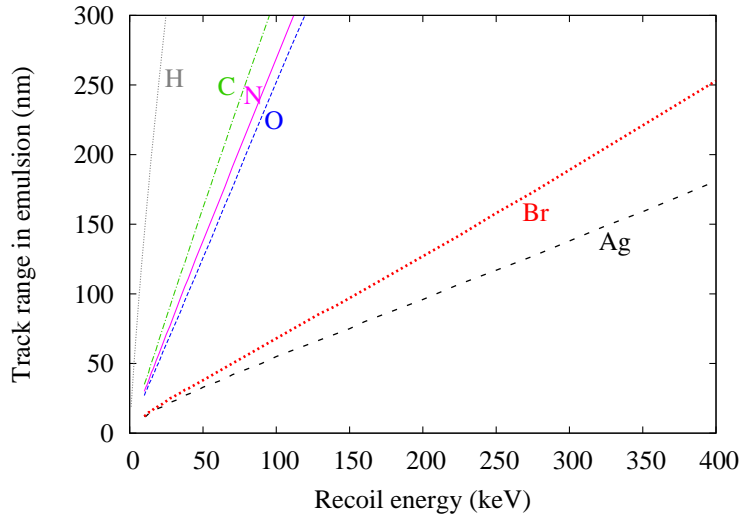}
\caption{Correspondence between ion track range in emulsion and the recoil energy.}
\end{center}
\end{figure}
For the nuclear emulsion, energy threshold corresponds to detectable range threshold of the nuclear emulsion. 
Light atoms such as C, N and O, have advantage to detect small recoil energy, i.e. light dark matter.
On the other hand, heavy ones including Ag and Br are not sensitive to light dark matter.
We converted recoil energy into track range by SRIM simulation \cite{SRIM}. 
The relationship between signal track range in matter and energy for each atom in nuclear emulsion is shown in Fig\ref{fig:correspondence}. 
The energy thresholds C, N and O are almost same as $33$ GeV which we adopt in the numerical calculations.

\section*{Appendix B: Concept of Background Rejection}
In the paper, the light dark matter mass region is focused. 
For realistic investigation, the rejection of back ground signals which can leave short track signal as well as the light dark matter on the nuclear emulsion should be discussed.
Basic concept of the back ground rejection is to create the nuclear emulsion which has no sensitivity for back ground signals by means of sensitivity control of nuclear emulsion itself and development treatment.
We shortly summarize the strategy of the background rejection though it may be too technical. 
\begin{itemize}
\item First we discuss how to deal with the background signals caused by radioactive sources from outside of the detector.
Size of signals can be adopted for background rejection because
the range of energy deposit per unit path length dE/dx is different between dark matter signals and radioactive backgrounds.
For example, expected total energy deposition for heavy targets and light targets by dark matter scattering are
about 1000-2000 keV/$\mu$m and 100-300 keV/$\mu$m, respectively. 
On the other hand, the energy deposit for electron and proton background are about 10 keV/$\mu$m and 50 keV/$\mu$m, respectively. 
Therefore, sensitivity optimization of the nuclear emulsion itself, 
and development treatment are essential to produce nuclear emulsion which is sensitive to only
the signals with high dE/dx.
\item Serious backgrounds are caused by internal sources in nuclear emulsion itself. 
Especially, $\beta$ or $\gamma$-rays
from ${}^{238}$Th chain, $\beta$-rays from ${}^{40}$K and ${}^{14}$C are expected. 
The dominant backgrounds among them are ${}^{40}$K
and ${}^{14}$C.
${}^{40}$K can be mixed in nuclear emulsion when it is produced from KBr.
By adopting NaBr instead of KBr to produce the nuclear emulsion, background from ${}^{40}$K can be avoided.
Finally serious backgrounds are ${}^{14}$C. 
In order to examine positive signal regions of other experiments,
the rejection power $\gsim10^9$ is required.
\item Neutron from rock in underground is also possible background source. Especially, the
recoiled protons induced by neutron with more than 7 keV kinetic energy can be background.
Neutron flux of this energy region is expected to be $10^{-6}$ /cm$^{2}$/sec \cite{Beli1989}. 
Supposing non-shielding and perfect detection efficiency, the event rate of proton recoil will be about 2.6$\times 10^3$ /kg/year.
In order to achieve zero background in 10 kg$\cdot$year exposure, rejection powers of 10$^{5-6}$ are required.
The rejection will be realized by neutron shield and sensitivity control. 
\item Finally, remained backgrounds can be discriminated by directionality because background tracks have angular distribution with isotropic. 
\end{itemize}








\end{document}